\documentclass[pra,aps,twocolumn]{revtex4}
\usepackage{amsmath}
\usepackage[dvips]{graphicx}
\usepackage{subfigure}

\def\cite#1{\buildlist{#1}\updaterefs{\listed}[{\citerefs{\listed}}]}

\def\quietcite#1{\buildlist{#1}\updaterefs{\listed}}
\newif\ifold
\def\buildlist#1{\oldcheck#1@%
                 \ifold\def\listed{#1}%
                 \else\def\listed{}\makelistmacro#1,@%
                 \fi}
\def\oldcheck#1#2@{\def\delim{#1}\def\olddelim{\\}%
                  \ifx\delim\olddelim%
                     \oldtrue\def\\{}%
                  \else %
                     \oldfalse %
                  \fi}
\def\makelistmacro#1#2,#3@{\add{#1#2}\to{\listed}%
                         \def\restoflist{#3}%
                         \ifx\restoflist\empty\relax%
                         \else\makelistmacro#3@%
                         \fi}
\def\refkeys{}
\def\biblist{}
\newcount\citenum
\newif\ifmember
\def\updaterefs#1{\citenum=0%
                 \def\\##1{\updatedef{##1}}%
                 #1}%
\def\updatedef#1{\advance\citenum by 1%
                 \ifx\refkeys\empty
                    \add{#1}\to{\refkeys}%
                 \else%
                    \memberfalse%
                    \ismember{#1}\of{\refkeys}
                    \def\\##1{\updatedef{##1}}
                    \ifmember
                    \else%
                       \add{#1}\to{\refkeys}
                    \fi%
                 \fi}%
\def\ismember#1\of#2{\def\given{#1}%
                     \def\\##1{\def\next{##1}%
                               \ifmember%
                               \else%
                                  \ifx\next\given%
                                      \membertrue%
                                  \fi%
                               \fi}%
                     #2}%
\toksdef\ta=0
\toksdef\tb=2
\long\def\add#1\to#2{\ta={\\{#1}}%
                     \tb=\expandafter{#2}%
                     \global\edef#2{\the\tb\the\ta}}%
\newcount\refnum
\newcount\seqnum
\newcount\seqfirst
\newcount\seqlast
\def\citerefs#1{\refnum=0%
                \seqnum=0%
                \def\\##1{\citedef{#1}{##1}}%
                \refkeys%
                \def\\##1{}}
\def\citedef#1#2{\ifnum\citenum>0
                     \advance\refnum by 1%
                     \memberfalse%
                     \ismember{#2}\of{#1}
                     \def\\##1{\citedef{#1}{##1}}
                     \ifmember
                        \advance\citenum by-1 %
                        \ifnum\seqnum=0 %
                           \ifnum\citenum=0 %
                              \number\refnum %
                           \else %
                              \advance\seqnum by 1 %
                              \seqfirst=\refnum %
                           \fi %
                        \else %
                           \advance\seqnum by 1 %
                           \ifnum\citenum=0 %
                              \ifnum\seqnum=2 %
                                 \number\seqfirst ,\number\refnum %
                              \else %
                                 \number\seqfirst \hbox{--}\number\refnum %
                              \fi %
                           \else %
                              \seqlast=\refnum %
                           \fi %
                        \fi %
                     \else %
                        \ifnum\seqnum>2 %
                           \number\seqfirst \hbox{--}\number\seqlast ,%
                        \else %
                           \ifnum\seqnum=2 %
                              \number\seqfirst ,\number\seqlast ,%
                           \fi %
                           \ifnum\seqnum=1 %
                              \number\seqfirst ,%
                           \fi%
                        \fi%
                        \seqnum=0 %
                     \fi %
                  \fi}
\long\def\listrefs{\refnum=0%
                   \def\\##1{\advance\refnum by 1%
                             \writeref{\number\refnum}{##1}}%
                   \refkeys}
\long\def\listrefsAIP{\refnum=0%
                      \parindent=0pt 
                      \def\\##1{\advance\refnum by 1%
                                \ifnum\refnum>8
                                   \parindent=0pt 
                                \fi
                                \writerefAIP{\number\refnum}{##1}}%
                      \refkeys%
                      \parindent=0pt}

\long\def\fookey{\refnum=0%
                   \def\\##1{\advance\refnum by 1%
                             \writekeyfile{\number\refnum}{##1}}%
                   \refkeys}
\long\def\writeref#1#2{\memberfalse%
                       \def\nextkey{#2}%
                       \long\def\\##1\\##2{\def\nextbibkey{##1}%
                                           \ifmember
                                           \else
                                              \ifx\nextkey\nextbibkey
                                                 \membertrue
                                                 \printref{#1}{##2}
                                              \fi
                                           \fi}%
                       \biblist%
                       \def\\##1{\advance\refnum by 1%
                                 \writeref{\number\refnum}{##1}}}
\long\def\writerefAIP#1#2{\memberfalse%
                          \def\nextkey{#2}%
                          \long\def\\##1\\##2{\def\nextbibkey{##1}%
                                              \ifmember
                                              \else
                                                 \ifx\nextkey\nextbibkey
                                                    \membertrue
                                                    \printrefAIP{#1}{##2}
                                                 \fi
                                              \fi}%
                          \biblist%
                          \def\\##1{\advance\refnum by 1%
                                   \ifnum\refnum>8
                                      \parindent=0pt%
                                   \fi
                                   \writerefAIP{\number\refnum}{##1}}}
\long\def\writekeyfile#1#2{\memberfalse%
                       \def\nextkey{#2}%
                       \long\def\\##1\\##2{\def\nextbibkey{##1}%
                                           \ifmember
                                           \else
                                              \ifx\nextkey\nextbibkey
                                                 \membertrue
                                                 \printkeyfile{#1}{##1}
                                              \fi
                                           \fi}%
                       \biblist%
                       \def\\##1{\advance\refnum by 1%
                                 \writekeyfile{\number\refnum}{##1}}}
\long\def\printref#1#2{\par\hang\indent\llap{{#1}. }\ignorespaces #2.}
\long\def\printrefAIP#1#2{\par[{#1}] \ #2.}
\long\def\printkeyfile#1#2{\write\bibfile{\string\quietcite{#2}}}
\def\bibitem#1#2{\add{#1}\to{\biblist}%
                 \add{#2}\to{\biblist}}
\begin{document}
\newcommand{\vc}{\mathbf}
\newcommand{\cV}{{\cal V}}
\newcommand{\vn}{\vec{\nabla}}
\newcommand{\be}{\begin{equation}}
\newcommand{\ee}{\end{equation}}
\newcommand{\vm}{{{\vec{m}}}}
\newcommand{\vM}{{{\vec{M}}}}
\newcommand{\vh}{{{\vec{h}}}}
\newcommand{\bk}{{{\bf{k}}}}
\newcommand{\bK}{{{\bf{K}}}}
\newcommand{\br}{{{\bf{r}}}}
\newcommand{\hbr}{{\hat{\bf{r}}}}
\newcommand{\hbR}{{\tilde{\bf{R}}}}
\newcommand{\bR}{{{\bf{R}}}}
\newcommand{\bq}{{\bf{q}}}
\newcommand{\bea}{\begin{eqnarray}}
\newcommand{\eea}{\end{eqnarray}}
\newcommand{\ra}{\rangle}
\newcommand{\la}{\langle}
\newcommand{\upa}{\uparrow}
\newcommand{\dna}{\downarrow}
\newcommand{\bS}{{\bf S}}
\newcommand{\vS}{\vec{S}}
\newcommand{\dg}{{\dagger}}
\newcommand{\pdg}{{\phantom\dagger}}

\title{Repulsive Fermi gas in a harmonic trap: Ferromagnetism and spin textures}
\author{L. J. LeBlanc$^1$, J. H. Thywissen$^1$, A. A. Burkov$^2$, A. Paramekanti$^1$}
\affiliation{$^1$ Department of Physics, University of Toronto, Toronto,
Ontario M5S1A7, Canada}
\affiliation{$^2$ Department of Physics, University of Waterloo, Waterloo,
Ontario N2L3G1, Canada}
\begin{abstract}
We study ferromagnetism in a repulsively 
interacting two-component Fermi gas in a harmonic trap.
Within a local density
approximation, the two components phase-separate beyond a
critical interaction strength, with one species
having a higher density at the trap center. We discuss several
easily observable experimental signatures of this transition.
The mean field
release energy, its separate kinetic and interaction contributions,
as well as the potential energy, all
depend on the interaction strength and contain a sharp 
signature of this transition. In addition, the conversion rate
of atoms to molecules, arising from
three-body collisions, peaks at an interaction strength just
beyond the ferromagnetic transition point.
We then go beyond
the local density approximation, and derive an
energy functional which includes a term that depends on the local
magnetization gradient and acts as
a `surface tension'. Using this energy functional, we numerically study 
the energetics of some candidate spin textures which may be 
stabilized in a harmonic trapping potential at zero net magnetization. 
We find that a hedgehog state has a lower 
energy than an `in-out' domain wall state
in an isotropic trap. Upon inclusion of trap anisotropy
we find that the hedgehog magnetization profile gets
distorted due to the surface tension term, this distortion being
more apparent for small atom numbers.
We estimate that the magnetic dipole interaction does
not play a significant role in this system. We consider possible implications  
for experiments on trapped $^6$Li and $^{40}$K gases.
\end{abstract}
\maketitle

\section{Introduction}
In recent years, a series of beautiful experiments have shown that a 
gas of two-component fermions interacting via a Feshbach resonance 
exhibits a superfluid state at low temperature 
\cite{greiner03,bourdel03,gupta03,zweirlein05}.
An exciting new direction for cold atoms experiments is the study of
ferromagnetism arising from repulsive interactions in a
two-component Fermi gas. 
Such a ferromagnetic `Stoner instability'  \cite{stoner33}
occurs, within mean field 
theory of a homogeneous Fermi gas at zero temperature, 
when the (repulsive) s-wave scattering length, $a_S$, between two 
spin states is large enough that $k_F a_S > \pi/2$, where 
$\hbar k_F$ is the Fermi momentum of the gas.
This condition can be satisfied upon tuning $a_S$
to large positive values near a Feshbach 
resonance provided the system stays stable for a sufficiently long
time.
Since ferromagnetism arises from two-body interactions, whereas
atom loss due to Feshbach molecule formation is because of three-body
collisions \cite{dpetrov03}, there may be
a range of densities where the lifetime is long enough to reach
the ferromagnetic state.

The suggestion that ferromagnetism may be achieved, as a metastable state,
in cold Fermi gases is not new. Salasnich and co-workers
\cite{salasnich00} have studied
the mean field theory of a harmonically trapped Fermi gas
with repulsive interactions and found that this should lead to phase
separation between the two spin species if the net magnetization is zero.
A similar study was carried out by Sogo and Yabu \cite{sogo02} allowing for
non-zero spontaneous magnetization.
Duine and MacDonald \cite{duine05} later 
showed that the ferromagnetic transition in
a homogeneous Fermi gas changes from a continuous to a first order 
transition at low enough temperatures upon going beyond
mean field theory. They also proposed that an initially magnetized Fermi
gas will tend to stay spin-coherent for long times, even in the presence of 
magnetic field
noise that is na\"ively expected to cause strong dephasing, provided
the system is close to the transition into a ferromagnetic state
\cite{duine05}.

Using an optical lattice and engineering the band structure to get
flat (dispersionless) bands is another 
interesting route to 
achieving ferromagnetism. Such `flat-band ferromagnetism' \cite{tasaki98}
has the advantage 
that the
ferromagnetic state occurs at weak repulsive interactions and can be 
theoretically analyzed in a reliable fashion;
however, an existing theoretical proposal along these lines 
involves working with fermions in the p-band of a honeycomb optical 
lattice \cite{congjun08} which is a significant experimental challenge.
A recent work \cite{coleman08} has considered the possibility of
ferromagnetism for strongly interacting fermions
in optical lattices and studied,
within a phenomenological Landau theory,
the energetics of possible spin textures (such as hedgehog states, 
domain walls and skyrmions) which might arise in a trap.

In this paper, motivated by earlier work and by ongoing experimental 
efforts, we revisit the problem of ferromagnetism in a harmonically
trapped two-component Fermi gas. 
We begin by using a ``local density approximation'' (LDA), sometimes
referred
to in the literature as the Thomas-Fermi approximation, to describe this
system. Within the LDA, 
we find that the mean field release energy of the trapped gas
(as well as the 
potential energy and the kinetic energy component of the release energy) 
provides a simple, albeit indirect, diagnostic of the 
ferromagnetic transition. We find that the formation of nonzero local
magnetization in the trap causes a suppression of the atom loss
rate via three-body collisions. This suppression competes with the growth 
of the
loss rate as the interactions get stronger, leading to a peak in the atom 
loss
rate at an interaction strength which is very close to, but slightly beyond, 
the ferromagnetic transition point.
 We then show how one might incorporate 
magnetization gradient (or `surface tension') terms 
in order to go beyond the LDA. Our energy functional
is akin to an earlier phenomenological Landau theory for ferromagnetism
in an optical lattice
\cite{coleman08} but explicitly keeps track of the spatial dependence of
all the Landau theory coefficients which arise from density variations
in the trap.
Using this extended energy functional, we  study the energetics of various
spin textures in the case where the net magnetization, which does not
relax in these quantum gases,
is assumed to be zero. This corresponds to choosing the
initial population to be the same for both hyperfine species of fermions.
We show, in this case, that a hedgehog configuration of the 
magnetization
has  a lower energy than an `in-out' phase-separated configuration with a
domain wall.
A similar phenomenon has been predicted for fermions in optical lattices 
\cite{coleman08}.
Finally, we turn to the effect of anisotropic trapping frequencies in
a harmonic trap. While such anisotropies can
be incorporated by a trivial rescaling of coordinates in the LDA,
this is no longer true in the presence of surface tension which leads to
a breakdown of the LDA. (A breakdown of the LDA has been observed 
\cite{hulet06}
and theoretically addressed  \cite{emueller,demler06,sensarma07}
 in the context of polarized superfluids 
in highly anisotropic traps.)
We use our
extended energy functional to numerically study how the hedgehog state 
distorts upon going from a spherically symmetric 
trap to an anisotropic cigar-shaped trap. 
We conclude with estimates which indicate that the magnetic
dipole interaction between atoms can be neglected for $^6$Li and
$^{40}$K.

\section{Ferromagnetism within the local density approximation}
The Hamiltonian describing a uniform two-component Fermi gas 
interacting through a repulsive s-wave contact interaction is given by
\be
H  \!=\! \sum_\sigma \!\int \!\frac{d^3\bK}{(2\pi)^3} \epsilon_\bK
c^\dg_{\bK\sigma} c^{\pdg}_{\bK\sigma} \!\!+ \!g \int \!\! d^3\bR
c^\dg_{\bR\upa}
c^\dg_{\bR\dna}
c^\pdg_{\bR\dna}
c^\pdg_{\bR\upa},
\ee
where $\epsilon_\bK = \hbar^2 \bK^2/2 M$ is the kinetic energy
of atoms with mass $M$ and momentum $\hbar\bK$.
For a Fermi gas with $N_\sigma$ particles of spin-$\sigma$,
the uniform gas densities of each spin is $\rho_\sigma=N_\sigma/\cV$,
and the total kinetic energy of the uniform gas is just
\be
K = \frac{3}{5} \cV (E_{F\upa} \rho_\upa + E_{F\dna} \rho_\dna),
\ee
where
$\cV$ denotes the system volume, and
$E_{F\sigma} =  \alpha \rho_\sigma^{2/3}$,
with $\alpha= (6 \pi^2)^{2/3} \hbar^2/2 M$,
denotes the Fermi energy of particles with spin-$\sigma$.
A mean field theory of the interacting Hamiltonian yields the
total interaction energy
\be
U = g \cV \rho_\upa \rho_\dna.
\ee
At this level of treatment the contact interaction strength
$g$ is related to the two-body scattering length $a_S$
in vacuum via 
$g=4\pi a_S \hbar^2/M$.

The local density approximation (LDA) for a trapped Fermi gas
corresponds to simply assuming that
the above results apply locally in the presence of a trap potential
$V(\bR)$.
The ground state energy of this trapped Fermi gas 
is then obtained by minimizing the
energy functional
\bea
E[\{\rho_\sigma(\bR)\}]
\!\! &=& \!\! \int \!\! d^3\bR \!\!\left[ \frac{3}{5} \alpha
\sum_\sigma \rho_\sigma^{5/3}(\bR)
\! + \!g \rho_\upa(\bR) \rho_\dna(\bR) \right. \nonumber \\
&+&\!\! \left. V(\bR) \sum_\sigma 
\rho_\sigma(\bR) - \sum_\sigma \mu_\sigma \rho_\sigma(\bR) \right],
\eea
where $\{\rho_\sigma(\bR)\}$ denotes the density profile of both spin
species
$[\rho_\upa(\bR),\rho_\dna(\bR)]$. Here we have introduced two Lagrange
multipliers $\mu_\sigma$ which act as chemical potentials for the two spin
species and serve to impose the constraints 
$\int d^3\bR \rho_\sigma(\bR) = N_\sigma$. The separate constraint on each
spin component arises from the assumption that the two spin components
correspond to the lowest two Zeeman split hyperfine levels of a Fermi gas.
Since the Zeeman splitting near a Feshbach resonance is typically
far greater than all other energy scales and the total energy must be conserved 
in these thermally isolated gases, we arrive at the constraint that
the population of the two Zeeman components
cannot change for fermionic atoms where the only interaction is between
different spin components. Thus,
unlike in solid state
ferromagnets, the magnetization can be conserved on very long time scales.

\subsection{Rescaling to the isotropic problem for a harmonic trap}
Let us assume that the trapping potential is harmonic, but possibly 
anisotropic, so 
that $V(\bR) = \frac{1}{2} M \sum_i \omega_i^2 \bR_i^2$. Here
$\omega_{x,y,z}$ are the trapping frequencies along different directions.
In order to make the problem appear isotropic we can rescale distances
by setting $\hbR_i = \bR_i (\omega_i/\Omega)$, where $\Omega = 
(\omega_x \omega_y \omega_z)^{1/3}$ is the geometric mean of the trap
frequencies. With this standard rescaling, we get
\bea
E[\{\rho_\sigma(\hbR)\}]
&\!\!=\!\!& \!\!\int \!\! d^3 \hbR \!\!\left[ \frac{3}{5} \alpha
\sum_\sigma \rho_\sigma^{5/3}(\hbR)
+ g \rho_\upa(\hbR) \rho_\dna(\hbR) \right. \nonumber \\
&+& \left. \!\! \frac{1}{2} M \Omega^2 \hbR^2 \sum_\sigma
\rho_\sigma(\hbR) \!-\! \sum_\sigma \mu_\sigma \rho_\sigma(\hbR) \right].
\eea

\subsection{Noninteracting unmagnetized gas}
For the unmagnetized gas, we have $N_\upa=N_\dna=N/2$ and for the 
noninteracting case we can set $g=0$. This reduces the energy
functional to
\bea
E^0_N[\{\rho_\sigma(\hbR)\}]\!\!&=&\!\!
\sum_\sigma \! E^0_{N\sigma}[\rho_\sigma(\hbR)], \\
E^0_{N\sigma}[\rho_\sigma(\hbR)] 
\!\!&=&\!\! \int\! d^3\hbR \! \left[ \frac{3}{5} \alpha
\rho_\sigma^{5/3}(\hbR)
\!+\! \frac{1}{2} M \Omega^2 \hbR^2 
\rho_\sigma(\hbR) \right. \nonumber\\
\!\!&-&\!\! \left. \mu_\sigma \rho_\sigma(\hbR) \right].
\eea
Setting $\delta E^0_N/\delta \rho_\sigma = 0$ leads to the equations
\be
\alpha \rho_\sigma^{2/3}(\hbR)
= (\mu^0_N- \frac{1}{2} M \Omega^2 \hbR^2),
\ee
where we have used symmetry to set $\mu_\upa=\mu_\dna=\mu^0_N$.
The solution to this equation is simply
\be
\rho_\sigma(\hbR) = \alpha^{-3/2} \left[\mu^0_N- \frac{1}{2} 
M \Omega^2 \hbR^2\right]^{3/2}.
\ee
Clearly there is a maximum radius,
\be
R^0_N =  \sqrt{\frac{2 \mu^0_N}{M \Omega^2}},
\ee
beyond which $\rho_\sigma(\hbR)=0$. Integrating upto this maximum radius
using spherical symmetry of the density, and employing the constraint,
we end up with
\bea
\mu^0_N &\equiv& E^0_F = \hbar \Omega (3 N)^{1/3}, \\
R^0_N &=&  \sqrt{\frac{2 E^0_F}{m_a \Omega^2}} = a_{\rm HO} (24 N)^{1/6},\\
E^0_N &=&  \frac{3}{4} N E^0_F = \frac{\hbar \Omega}{4} (3 N)^{4/3}, \\
\rho^0_{N,\sigma}(0)&=& \frac{4 N}{\pi^2 (R^0_N)^3} =
a_{\rm HO}^{-3} (\frac{2}{3\pi^4})^{1/2} N^{1/2},
\eea
where $a_{\rm HO} = (\hbar/M\Omega)^{1/2}$ is the oscillator length,
and we denote the density solution for this noninteracting unmagnetized
Fermi gas by $\rho^0_{N\sigma}(\hbR)$. Here $\mu^0_N$ is the chemical
potential of the gas, $E^0_N$ is the total energy of the gas,
and $\rho^0_N(0)$ is the atom density at the trap center.

\subsection{Converting the interacting problem to dimensionless variables}
Let us use the noninteracting unmagnetized gas results to convert 
to dimensionless variables as follows.
\bea
\br &=& \frac{\hbR}{R^0_N}, \\
n_\sigma &=& \frac{\rho_\sigma}{\rho^0_{N\sigma}(\hbR=0)}, \\
\lambda &=& k^0_F(0) a_S, \\
\varepsilon &=& E/E^0_N, \\
\gamma_\sigma &=& \mu_\sigma/\mu^0_N.
\eea
Here $\lambda$ is the dimensionless interaction parameter,
$\varepsilon,\gamma_\sigma$ are the total energy and chemical potential
respectively in dimensionless units,
and $k^0_F(0)$ denotes the Fermi wavevector at the
trap center for the unmagnetized noninteracting gas. 

In terms of these dimensionless variables, the energy functional
becomes
\bea
\varepsilon[\{n_\sigma(\br)\}]
\!\!&=&\!\! \frac{16}{3\pi^2}
\int d^3 \br \left[  \frac{3}{5} (n_\upa^{5/3}(\br) + n_\dna^{5/3}(\br)) \right.\nonumber\\
&\!+\!&\!\!\!\!\! \left. \frac{4 \lambda}{3\pi} n_\upa(\br) n_\dna(\br) \!\! - \!\!
\sum_\sigma 
(\gamma_\sigma \!\!-\!\! \br^2) n_\sigma(\br) \right].
\label{eq:LDA}
\eea
If we assume that the ground state solution for the densities 
respects the spherical symmetry of this energy functional, we can
further simplify the energy functional to a one-dimensional integral
\bea
\varepsilon[\{n_\sigma(r)\}]
\!\!&=&\!\! \frac{64}{3\pi}
\int d r ~ r^2 \left[  \frac{3}{5} (n_\upa^{5/3}(r) + n_\dna^{5/3}(r)) \right.\nonumber\\
&\!+\!&\!\!\!\!\! \left. \frac{4 \lambda}{3\pi} n_\upa(r) n_\dna(r) \!\! - \!\!
\sum_\sigma 
(\gamma_\sigma \!\!-\!\! r^2) n_\sigma(r) \right].
\eea

\subsection{Variational minimization}
The variational minimization $\delta E / \delta n_\sigma(r)$ leads to
the following two equations
\bea
n_\upa(r)&=& \left[ (\gamma_\upa - r^2- \frac{4}{3\pi} \lambda n_\dna(r)) 
\right]^{3/2}, \\
n_\dna(r)&=& \left[ (\gamma_\dna - r^2- \frac{4}{3\pi} \lambda n_\upa(r)) 
\right]^{3/2},
\eea
subject to the constraints
\be
4\pi \int dr ~ r^2 n_\sigma(r) = \frac{\pi^2}{4} \frac{N_\sigma}{N}.
\ee
These equations can be iteratively solved (numerically) for the fermion 
densities given the interaction 
strength and the total fermion numbers for each species. 
Having solved them
we can use the resulting fermion densities
to compute physical observables. We will denote the
average magnetization by $\bar{m}
\equiv (N_\upa-N_\dna)/(N_\upa+N_\dna)$. For an
unmagnetized gas, we find that increasing the interaction $\lambda$
progressively modifies the density profile of the gas from that of
a noninteracting Thomas-Fermi profile at $\lambda=0$ to that of a 
fully polarized gas when $\lambda \gg 1$.

\subsection{Release energy and `ferromagnetic transition'}

\begin{figure}
\includegraphics[width=2.5in]{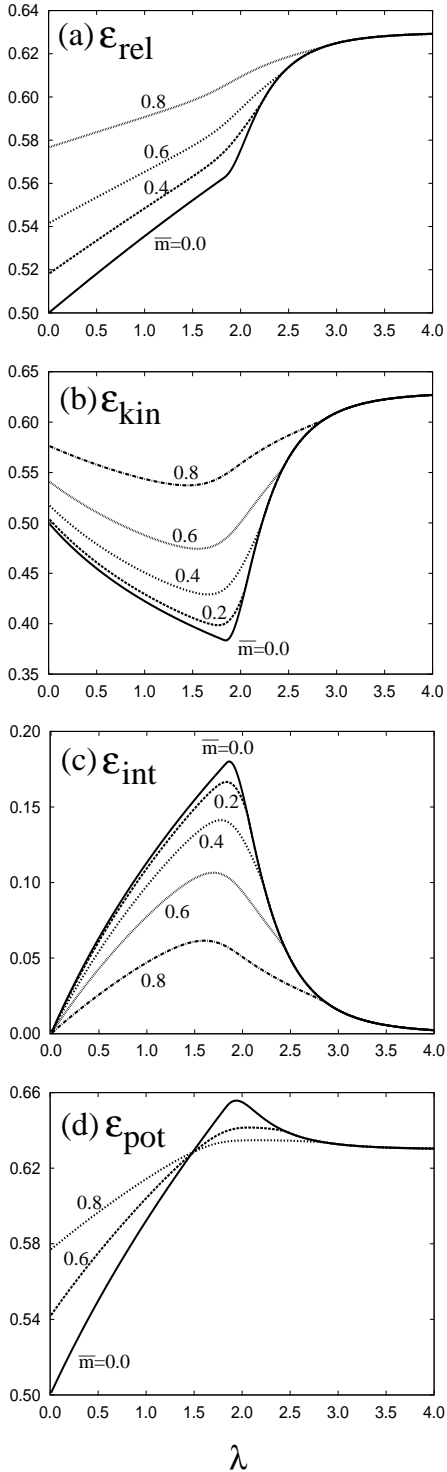}
\caption{(a): Dimensionless mean field release energy, $\varepsilon_{\rm rel}$, versus
interaction, $\lambda$, within the LDA
for indicated values of
$\bar{m}=(N_\upa-N_\dna)/N$. 
For $\bar{m}=0$, there is a
phase separation transition which
appears as a sharp kink in the release energy at $\lambda\approx 1.84$.
(b,c,d): Same as in (a) but for the kinetic energy, interaction energy and
potential energy of the gas.
The energy per particle in physical units
may be obtained by multiplying these results by $3 E^0_F/4$ where 
$E^0_F=\hbar\Omega(3N)^{1/3}$. As shown later,
going beyond the LDA
leads to negligible quantitative corrections to these results.}
\label{fig:release}
\end{figure}

The release energy of the trapped atomic gas
is measured by rapidly switching off the trap potential
and measuring the total kinetic energy of the atoms after some
time delay. Assuming that the switch-off process is instantaneous and
that all the interaction energy in the
initial state has been converted into the kinetic energy of atoms at the time of
measurement, the release energy and its separate kinetic and interaction
energy components are given, within the LDA, by
\bea
\varepsilon_{\rm rel} \!\! &=& \!\!
\frac{64}{3\pi} \!
\int \! \!dr~r^2 \!\!\left[ \frac{3}{5} \sum_\sigma
n_\sigma^{5/3}(r) \!+\! \frac{4}{3\pi} \lambda n_\upa(r) n_\dna(r)\! \right]\!\!, \\
\varepsilon_{\rm int} \!\! &=& \!\!
\frac{64}{3\pi} \!
\int \! \! dr~r^2 \!\!\left[\frac{4}{3\pi} \lambda n_\upa(r) n_\dna(r) \right], \\
\varepsilon_{\rm kin} \!\! &=& \!\!
\frac{64}{3\pi} \!
\int \!\! dr~r^2 \!\!\left[ \frac{3}{5} \sum_\sigma
n_\sigma^{5/3}(r) \right] .
\eea
The potential energy of the cloud, due to the confining harmonic trap,
can be easily obtained from experimental measurements of the cloud
profile and it is given by
\bea
\varepsilon_{\rm pot} \!\! &=& \!\!
\frac{64}{3\pi} \!
\int \! \! dr~r^2 \!\!\left[r^2 (n_\upa(r) + n_\dna(r)) \right].
\eea

From Fig.~\ref{fig:release}, we see that the release energy displays
a sharp transition point, for $\bar{m}=0$, at  $\lambda_{\rm crit} \approx 
1.84$. At this interaction strength, we find that
$k_F a_S = \pi/2$ at the trap center, with
$k_F$ being the Fermi wave vector at the trap center in the interacting
cloud, which corresponds to the Stoner transition point in the uniform
gas. Further, an examination of the density profile of the two
spin species shows that this corresponds to an onset of phase separation
in the trap ---  for $\lambda > \lambda_{\rm crit}$, atoms of one spin 
type tend to have a higher density at the trap
center while atoms of the other spin type are pushed away from the center
leading to a nonzero magnetization density near the trap center.
Exactly which atoms tends to accumulate at the center is a spontaneously
broken symmetry at zero magnetization, and this phase separation is simply
a local manifestation of ferromagnetic ordering.
This result for the
$\lambda_{\rm crit}$ at $\bar{m}=0$ translates into an estimate for
the critical two-body scattering length,
\be
a_S^{(\rm crit)} \approx 0.6 \lambda_{\rm crit} a_{\rm HO} N^{-1/6}
\approx a_{\rm HO} N^{-1/6},
\ee
beyond which one expects to see phase separation in the trap. For 
$\Omega/2\pi \approx 170$Hz we estimate for $N=10^4, 10^5, 10^6$, 
the respective critical scattering lengths
\bea
a_S^{(\rm crit)} (^{40}{\rm K}) &\approx& 
5300 a_0, 3600 a_0, 2500 a_0, \\
a_S^{(\rm crit)}(^6 {\rm Li}) &\approx& 
13800 a_0, 9500 a_0, 6400 a_0,
\eea
where $a_0 \approx 0.529\AA$ is the Bohr radius. 

Fig.~\ref{fig:release} also
shows that the kinetic energy and the interaction energy components of the
total release energy. Each of these observables shows a large and much more
dramatic signature at the transition (for $\bar{m}=0$) than the total release
energy. It would be promising to look for this signature in experiments. In addition,
the potential energy of the confined cloud also shows a maximum at the
ferromagnetic transition point.

Strictly speaking, there is no
phase transition (beyond mean field theory)
except in the thermodynamic limit which, for a trapped Fermi gas,
is obtained by taking $N \!\!\to \!\!\infty$ and $\Omega \!\! \to \! 0$ with $N \Omega^3$
held fixed.
For nonzero magnetization, however, there is no phase transition
even at mean field level; nevertheless, the release energy does display a 
fairly sharp 
crossover at $\lambda_{\rm crit}$ for $m \lesssim 0.2$.
The measured
release energy can only tell us about the existence of a phase transition --- 
for $\lambda > \lambda_{\rm crit}$, {\it in situ} 
measurements of the magnetization profile, which we discuss below, are
needed  to show that this
transition corresponds to ferromagnetism in the trap.

\subsection{Atom loss rate}
\begin{figure}
\includegraphics[width=2.8in]{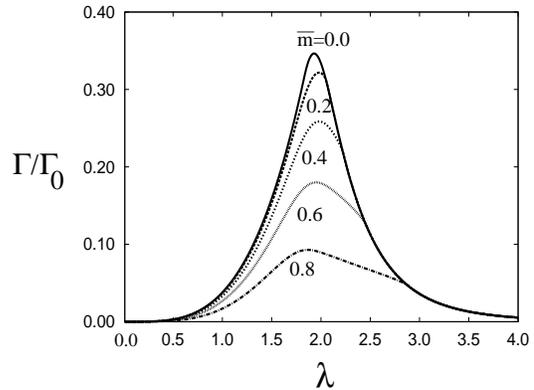}
\caption{Dimensionless atom loss rate, $\Gamma/\Gamma_0$, as a function
of interaction strength at various average magnetizations.}
\label{fig:loss}
\end{figure}

Atoms on the repulsive side of the Feshbach resonance tend to be unstable
to formation of molecules via three-body collisions. Apart from kinematic and
statistical contraints on these processes, there is a simple constraint that 
two of these atoms, which
eventually form the molecule, must have opposite spins.
One consequence of having a nonzero local magnetization in the trapped
gas is a suppression of the probability of finding fermions with opposite spin
in the same region of space, which leads to a strong suppression of such
three-body losses. A measured drop in the atom loss rate as a function of
increasing interaction strength would thus hint at the
presence of nonzero local magnetization in the trap. Upto an unknown
prefactor, $\Gamma_0$, we can
estimate this three-body loss rate as
\be
\Gamma =\Gamma_0 \lambda^4 
\int d^3\br~n_\upa(\br) n_\dna(\br) (n_\upa(\br)+n_\dna(\br)),
\ee
where the $\lambda^4$ scaling follows from a study of the three-fermion
problem \cite{dpetrov03}. Fig.~\ref{fig:loss} depicts a plot of $\Gamma/\Gamma_0$
as a function of the interaction strength $\lambda$. For $\bar{m}=0$,
the very rapid growth of
$\Gamma/\Gamma_0$ for small interaction strength arises from the rapidly
growth of the $\lambda^4$ coefficient,
while the rapid drop beyond the ferromagnetic transition point arises
from the formation of a nonzero magnetization which suppresses the
product $n_\upa(\br) n_\dna(\br)$ in the integrand. These two competing effects
lead to a peak in the rate of atom loss,
via conversion to molecules, at an interaction strength which is slightly beyond
the ferromagnetic transition point.

\section{Beyond the LDA: Magnetization gradients}

The discussion in the preceding section has focused on the properties of the
Fermi gas within the LDA. The energy functional at this level of approximation does
not have any gradient terms. We will not worry about the shortcomings of this approximation
for the density profile --- it is well known that the LDA breaks down near the trap edges ---
but instead focus on going beyond the LDA by considering magnetization gradient terms
with a view to studying the energetics of spin textures.
We begin by noting that although we have been assigning
a global spin axis to the magnetization, the LDA energy functional would be unchanged
if we in fact choose the local spin quantization axis to vary from point to point; only
the magnitude of the local magnetization plays a role. In
order to go beyond the LDA and to study the energies of various 
spin textures
in such a Fermi gas, we therefore need to extend the energy functional in two respects.
First, we have to promote the local magnetization to a vector quantity so the
magnetization can point in different directions on the Bloch sphere at different
spatial
locations. Second, we have to include terms in the energy functional which
depend on the local magnetization gradients; this
corresponds to adding a `surface tension'
term to the energy functional. 
The results from such an extended energy functional
should be compared, in the future, with microscopic
Hartree-Fock calculations.

We start with the dimensionless energy functional in Eq.(\ref{eq:LDA}) and
set 
\bea
n_\upa(\br) &=& \frac{n(\br)}{2} (1+m(\br)), \\
n_\dna(\br) &=& \frac{n(\br)}{2} (1-m(\br)),
\eea
which defines the local magnetization density $m(\br)$. As discussed,
the spin quantization axis can be chosen to be different at each space point
within the LDA.
Let us next expand the energy functional in powers of $m(\br)$; we will
keep terms upto $m^6(\br)$ although terminating the
expansion at $m^4(\br)$ would not qualitatively affect our results. 
The energy functional then
splits into two parts as
\be
\varepsilon=\varepsilon_a[n(\br)] + \varepsilon_b[n(\br),m(\br)],
\ee
where
\bea
\varepsilon_a[n(\br)]
\!\!&=&\!\! \frac{16}{3\pi^2}
\int d^3 \br \left[  \frac{6}{5} (\frac{n(\br)}{2})^{5/3}
+ \frac{\lambda}{3\pi} n^2(\br) \right.\nonumber\\
&-& \left. (\gamma-\br^2) n(\br) \right],
\label{eq:rhofun} \\
\varepsilon_b[n(\br),m(\br)]
&=& \frac{16}{3\pi^2} \!\!
\int \!\! d^3\br \!\! \left[
A_2(\br) m^2(\br) \right.\nonumber\\
&+&  
A_4(\br)
m^4(\br) +
A_6(\br)
 m^6(\br)  \nonumber \\
&-& \left. h~n(\br) m(\br) \right].
\label{eq:magfun}
\eea
Here $\varepsilon_a$ only depends on the density profile which depends on the interaction
$\lambda$ and which we assume is unchanged from
that given by the LDA calculation earlier. This is a good approximation since
the corrections to the LDA
energy are weak for typical atom numbers used in experiments as we will see below. 
The coefficients of the 
magnetization-dependent energy functional, $\varepsilon_b$, are
\bea
A_2(\br)&=&(\frac{n^{5/3}(\br)}{2^{2/3} 3} - \frac{\lambda}{3\pi} n^2(\br)), \\
A_4(\br)&=&\frac{n^{5/3}(\br)}{2^{2/3} 81}, \\
A_6(\br)&=&\frac{7 n^{5/3}(\br)}{2^{2/3} 2187}.
\eea
$A_2(\br)$
depends on $\lambda$ explicitly. In addition, all the coefficients $A_{2,4,6}$
depend on the spatial location in the trap through the density, and thus also depend 
implicitly
on the interaction strength $\lambda$. This dependence was
ignored in earlier phenomenological work on trapped fermions in an optical lattice
\cite{coleman08}.
The Lagrange multiplier in the energy functionals are given by
$\gamma = (\gamma_\upa+\gamma_\dna)/2$ and $h=(\gamma_\upa-\gamma_\dna)/2$.
Promoting the 
magnetization and the Lagrange multiplier $h$ to vectors $\vm,\vec{h}$,
and including gradient terms leads to an energy functional of the form
to
\bea
\varepsilon_b[n(\br),\vm(\br)]
\!\!&=& \!\!\!\!\frac{16}{3\pi^2}\!
\int \!\!d^3 \br \! \left[
A_2(\br) |\vm(\br)|^2 \right. 
\nonumber\\
\!\!&\!\!+\!\!&\!\!\!\! A_4(\br) |\vm(\br)|^4
+  A_6(\br) |\vm(\br)|^6 \nonumber \\
\!\!&\!\!+\!\!&\!\!\!\! \left. \frac{1}{2} \zeta_s(\br) 
\alpha_i (\nabla_i m_j(\br))^2
\!\!- \!\! \vh(\br)\!\cdot\!\vm(\br) \!\right]\!\!.
\eea
Here, $\alpha_i = (\omega_i/\Omega)^2$ which comes from our
rescaling to an isotropic problem.
The stiffness $\zeta_s(\br)$ depends on $\br$ only through
the density $n(\br)$, and it can be computed in the uniform Fermi 
gas assuming that the magnetization variation is slow on the scale of
the interparticle spacing, but fast on the length scale over which the
total density varies, so that density variations can be ignored in
this computation.
The Lagrange multiplier $\vh(\br)$ must be chosen to satisfy global
constraints on the magnetization, for instance,
$\int d^3\br~n(\br) m_i(\br) = 0$ for each component $i$. We next outline
the derivation of the stiffness term.

\subsection{Computation of the stiffness $\zeta_s(\br)$}
For small magnetization, we can obtain the result for $\zeta_s$ from
the result for the magnetic susceptibility of the uniform Fermi gas. 
Note that the excess
energy in an applied field $\vec{h}(\bq)$ (pointing in any direction) is
given by $\Delta E(\bq) = \frac{1}{2} \chi(\bq) h_i(\bq) h_i(-\bq)$ which
defines the wavevector dependent magnetic susceptibility. This tells us
that the magnetization $M(\bq)$ in this external field is simply
$M(\bq) = \chi(\bq) h(\bq)$, so that we can instead set
$\Delta E(\bq) = \frac{1}{2} \chi^{-1}(\bq) M_i(\bq) M_i(-\bq)$.
Expanding $\chi^{-1}(\bq) = \chi^{-1}_0 (1 + b \bq^2)$ then yields
\be
\Delta E(\bq) = \frac{1}{2} \chi^{-1}_0 (1+b \bq^2) M_i(\bq) M_i(-\bq).
\ee
The well-known result for a Fermi gas at $T=0$ is that $b=1/12
k_F^{2}$, using which the energy cost becomes, in real space, 
\be
\Delta E \!= \!\frac{1}{2\chi_0}\!\!\int \! d^3 \bR \! \left[
|\vM(\bR)|^2 \!+\! \frac{1}{12 k^2_F} (\vn M_i(\bR))^2
\right],
\ee
where
\be
\chi^{-1}_0 = \frac{\pi^2 \hbar^2}{M k_F} = \frac{\pi^2\hbar^2}{M}
(3\pi^2\rho)^{-1/3}.
\ee
Rescaling distances to get an isotropic harmonic trapping potential,
and setting $M_i = \rho^0_{N\sigma}(\br=0) n(\br) m_i(\br)$,
with $r = R/R_N^0$, we find
\bea
\zeta_s(\br) &=& \frac{n^{-1/3}(\br)}{2^{2/3} 3} 
\frac{1}{6 (3\pi^2n(\br))^{2/3}}
\left(\frac{1}{\rho_{N\sigma}^0 (R_N^0)^3} \right)^{2/3} \nonumber \\
&=& \frac{1}{72 n(\br) (3 N)^{2/3}}.
\label{eq:stiffness}
\eea
For general values of the magnetization, higher order gradient terms might
also become important. We will focus here on the effects of this simplest
gradient term in the energy functional.

\subsection{Simplified magnetization energy functional}
Before proceeding to the energetics of various spin textures, let us slightly
simplify the energy functional. Notice
that $n(\br)$ varies over the length scale of $1$ in our dimensionless units.
For large atom numbers, the stiffness is small as seen from Eq.(\ref{eq:stiffness})
and we therefore
expect significant variations of the magnetization to occur on length scales
$\ell \ll 1$ in our dimensionless units.
Making this assumption,
we can set
$\vn (n(\br) m_i(\br)) \approx n(\br) \vn m_i(\br)$, which results in the
slightly simplified
energy functional
\bea
\varepsilon_b
&=& \frac{16}{3\pi^2}
\int d^3 \br \left[
A_2(\br) |\vm(\br)|^2 
+ A_4(\br) |\vm(\br)|^4 \right.\nonumber\\
&+& A_6(\br) |\vm(\br)|^6 
+ \vec h(\br)\cdot\vec m(\br)\nonumber \\
&+& \left. \frac{n(\br)}{144 (3 N)^{2/3}} \alpha_i
(\nabla_i m_j(\br)) (\nabla_i m_j(\br))
\right],
\eea
where $\vec h(\br)$ is chosen to satisfy
\be
\int d^3 \br  ~n(\br)
m_i(\br)=0,
\ee
for each
component $i$ (for zero net magnetization).
Recall that $\alpha_i = (\omega_i/\Omega)^2$, where
$\Omega
=(\omega_x \omega_y \omega_z)^{1/3}$ is the geometric mean of the trap
frequencies.

\section{Energetics of spin textures}
The energy functional we have derived above allows us to study the
energetics of various magnetization patterns in the trapped 
Fermi gas. We begin by considering the
isotropic harmonic trap, for which we compare energies of a
hedgehog configuration and a domain wall configuration
of the magnetization. We then turn to an anisotropic cigar-shaped
trap and show how the hedgehog state  gets deformed from the
isotropic case. In each case, we begin by constructing the
appropriate ansatz for the magnetization. We then numerically
minimize the resulting energy functional, by discretizing it on a fine
grid of points and using a simulated annealing
procedure, to obtain the optimal magnetization profile and its energy.
Assuming that the density and magnetization satisfy the constraints
that the total atom number is fixed and the total magnetization is zero
(so that the Lagrange multipliers can be dropped),
we can express the total energy as a sum $\bar{\varepsilon}
=\varepsilon_1 + \varepsilon_2$ where
\bea
\varepsilon_1
\!\!&=&\!\! \frac{16}{3\pi^2}
\int \!\!d^3\br \left[  \frac{6}{5} (\frac{n(\br)}{2})^{5/3}
\!\!+\!\! \frac{\lambda n^2(\br)}{3\pi} \!+\! r^2 n(\br) \right], \\
\varepsilon_2&=&
\frac{16}{3\pi^2}
\int d^3 \br \left[
A_2(\br) |\vm(\br)|^2 
+ A_4(\br) |\vm(\br)|^4 \right.\nonumber\\
&+& \left. A_6(\br) |\vm(\br)|^6
\!+\! \frac{n(\br)}{144 (3 N)^{2/3}} \alpha_i
(\nabla_i {\vec m}_j(\br))^2
\right]\!\!.
\eea

\subsection{Isotropic trap: Hedgehog state}
For the isotropic trap, the density profile is spherically
symmetric, which allows us to set
\bea
\varepsilon_1
\!\!&=&\!\! \frac{64}{3\pi}
\int \!\!dr~r^2 \left[  \frac{6}{5} (\frac{n(r)}{2})^{5/3}
\!\!+\!\! \frac{\lambda n^2(r)}{3\pi} \!+\! r^2 n(r) \right].
\eea
For the magnetization-dependent energy functional, we must set 
$\alpha_i=1$ in the isotropic trap, and the hedgehog state
corresponds to choosing $\vec m (\br) = m(r) \hat{r}$. This leads to
\bea
\varepsilon_2
&=& \frac{64}{3\pi}
\int dr~r^2 \left[
A_2(r) m^2(r) \right.\nonumber\\
&+& A_4(r) m^4(r) + A_6(r) m^6(r)\nonumber\\
&+& \left. \frac{n(r)}{144 (3 N)^{2/3}}
\left\{2 \frac{m^2(r)}{r^2} + \left(\frac{d m(r)}{d r}\right)^2 \right\}
\right].
\label{eq:hh}
\eea
We do not have to pay attention to the
zero magnetization constraint since this is guaranteed for any
choice of $m(r)$ by the hedgehog ansatz symmetry. For typical particle
numbers in experiments, $N \sim 10^4-10^6$, the stiffness term has a very
small coefficient. We will therefore assume that the average
density profile $n(r)$ obtained
from our earlier LDA calculation remains unchanged and only focus on
changes in the magnetization profile arising from inclusion of gradient terms.

\begin{figure}
\includegraphics[width=3.5in]{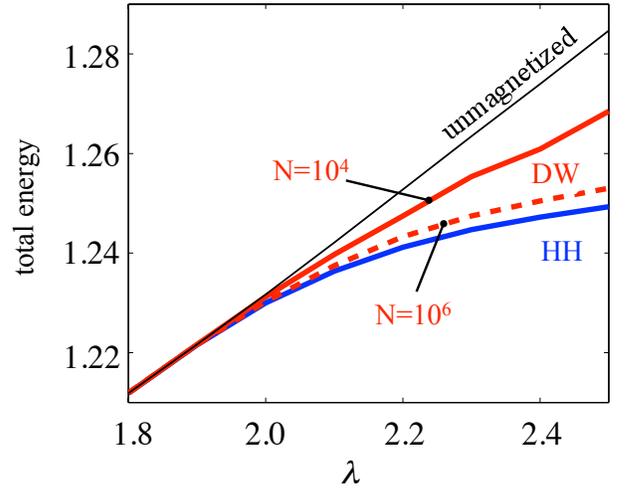}
\caption{(Color online) Dimensionless total 
energy, $\varepsilon_1+\varepsilon_2$, shown as a function of
interaction strength, $\lambda$, for an isotropic harmonic trap. 
DW indicates the energy of the
domain wall state for $10^4$ atoms (solid) and $10^6$ atoms (dashed).
HH denotes $\varepsilon_1+\varepsilon_2$ for the hedgehog state which is 
nearly identical for $10^4$ and $10^6$ atoms.
Also shown (thin solid line, `unmagnetized') is
$\varepsilon_1$, defined in Eq.(44),
which depends only on the total density profile.}
\label{fig:isoenergy}
\end{figure}

Fig.~\ref{fig:isoenergy} shows the energy $\bar{\varepsilon}^{HH}$ 
of the hedgehog state obtained by 
finding the function $m(r)$ which minimizes the hedgehog state energy. 
Fig.~\ref{fig:isomag}(a)
shows the magnetization profile of the hedgehog state at two different interaction strengths.
We find that the magnetization is suppressed in a small region around the trap center
and vanishes at $r=0$. To understand the magnetization profile of the hedgehog near
the trap center,
we can focus just on the last two terms in Eq.(\ref{eq:hh}). Taking a
functional derivative with respect to $m(r)$ and setting it to
zero then suggests that $m(r) \sim r^2$ at small $r$, so the
energy density coming from the central region of the hedgehog is 
finite. Far from the center, we expect the magnetization to be small. These expectations
are consistent with the magnetization profiles shown in
Fig.~\ref{fig:isomag}(a).

\subsection{Isotropic trap: Domain wall state}
For the domain wall state we have, as before,
\bea
\varepsilon_1
\!\!&=&\!\! \frac{64}{3\pi}
\int \!\!dr~r^2 \left[  \frac{6}{5} (\frac{n(r)}{2})^{5/3}
\!\!+\!\! \frac{\lambda n^2(r)}{3\pi} \!+\! r^2 n(r) \right].
\eea
For the magnetization dependent energy functional, we set
$\alpha_i=1$ in the isotropic trap
and choose $\vm(\br) = m(r) \hat{z}$ for the domain wall. This is capable of
describing a state with spin-$\upa$, say, at the trap center with
spin-$\dna$ pushed away from the center, what we might call an
`in-out' domain wall. We find
\bea
\varepsilon_2
&=& \frac{64}{3\pi}
\int dr~r^2 \left[
A_2(r) m^2(r) + A_4(r) m^4(r) \right.\nonumber\\
\!&\!+\!&\! \left. A_6(r) m^6(r)\!+\!
\frac{n(r)}{144 (3 N)^{2/3}}
\left(\!\frac{d m(r)}{d r}\!\right)^2
\right]\!\!.
\eea
where, for $N_\upa=N_\dna$, we must satisfy the constraint
$\int dr~r^2 n(r) m(r)=0$.
\begin{figure}
\includegraphics[width=3.5in]{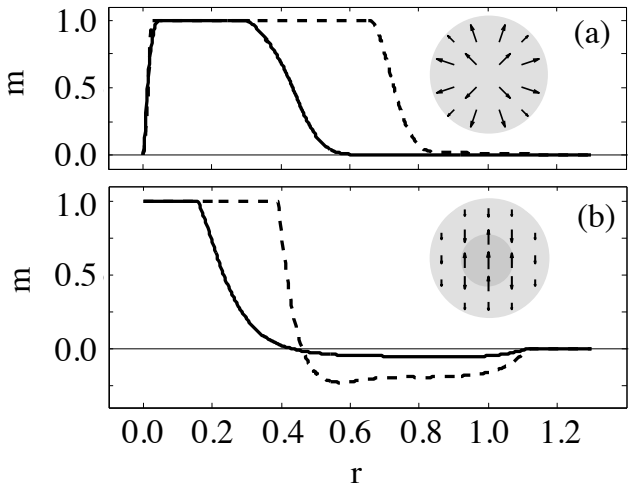}
\caption{(a) Magnetization profiles for the hedgehog state 
at $\lambda = 2.0$ (solid) and $\lambda = 2.4$ (dashed). (b) Magnetisation profiles 
for the domain wall state at $\lambda = 2.0$ (solid)  and $\lambda = 2.4$ (dashed).  
The profiles have been calculated for $10^4$ atoms in an isotropic trap.
The hedgehog state has zero magnetization at the trap
center while the domain wall state magnetization gets 
suppressed around the domain wall but remains nonzero at the trap center. Insets
indicate the schematic magnetization plot of the hedgehog state and the domain wall state.}
\label{fig:isomag}
\end{figure}
Fig.~\ref{fig:isoenergy} shows the energy $\bar{\varepsilon}^{DW}$ 
of the domain wall
state obtained by 
finding the function $m(r)$ which minimizes its energy subject to the 
zero magnetization
constraint. We find that this domain wall state has a higher energy than the 
hedgehog state.
Fig.~\ref{fig:isomag}(b)
shows the magnetization profile of the domain wall state.
As expected, the magnetization is suppressed in a small region around the domain wall
but remains nonzero at the trap center.

\subsection{Cigar-shaped trap: Distorted hedgehog}
If we consider a cylindrically symmetric (cigar-shaped) trap, 
we can look at an ansatz of the form
\be
\vec{m}({\bf r}) = m(\rho,z) \left(\frac{x}{\rho} \sin\psi, \frac{y}{\rho}
\sin\psi, \cos\psi \right),
\label{eq:dist_hh}
\ee
where $\rho=\sqrt{x^2+y^2}$ and $\psi\equiv\psi(\rho,z)$.
For $\psi=\theta=\tan^{-1}(\rho/z)$ this reduces to the spherical 
hedgehog ansatz. 
Note that the direction of the 
magnetization on the Bloch sphere is unrelated to the location in
real space. We could equally well have chosen, for instance,
\be
\vec{m}({\bf r}) = m(\rho,z) \left(\cos\psi,\frac{x}{\rho} \sin\psi,\frac{y}{\rho}
\sin\psi\right).
\ee
With the choice of magnetization in Eq.(\ref{eq:dist_hh}), 
we have $|\vec m (\br)|^2 = m^2(\rho,z)$, while
\bea
\alpha_i (\partial_i m_j) (\partial_i  m_j) 
\!\!\!&=&\!\!\! \alpha_\perp (\partial_\rho m)^2 \!+\! \alpha_z
(\partial_z m)^2 \!+\!
\alpha_\perp \frac{m^2}{\rho^2} \sin^2\psi \nonumber\\
&+& m^2 \left[\alpha_\perp (\partial_\rho \psi)^2 + \alpha_z (\partial_z \psi)^2 \right],
\eea
so the integral $\int d^3 \br \to 2\pi \int dz d\rho \rho$. 
We can assume that $m$ is an even function of $z$
and that $\psi(\rho,-z)=\pi-\psi(\rho,z)$ (so that $\sin^2\psi(\rho,-z) = \sin^2\psi(\rho,z)$)
to restrict the energy integration grid to just $z > 0$. These conditions ensure that the
total magnetization integrates to zero.
The final expression for the energy can thus be recast, with $r\equiv\sqrt{\rho^2+z^2}$,
as
\bea
\varepsilon_1
\!\!&=&\!\! \frac{64}{3\pi}\!
\int \!\!\!dr~r^2\!\! \left[  \frac{6}{5} (\frac{n(r)}{2})^{5/3}
\!\!+\!\! \frac{\lambda n^2(r)}{3\pi} \!+\! r^2 n(r) \!\right] \\
\varepsilon_2
&=& \frac{64}{3\pi} \!\!\int_0^{R_{max}}\!\! dz~\int_0^{\sqrt{r^2_{max}-z^2}} \!\!\!\!d\rho~\rho~F(\rho,z) \\
F(\rho,z)&=&
A_2 m^2
+ A_4 m^4
+A_6 m^6 \nonumber\\
&+& \frac{n}{144 (3 N)^{2/3}} \left\{\alpha_\perp \frac{m^2}{\rho^2} \sin^2\psi \right.\nonumber\\
&+& \alpha_\perp (\partial_\rho m)^2+ \alpha_z (\partial_z m)^2 \nonumber \\
&+& \left. m^2 \left[\alpha_\perp (\partial_\rho \psi)^2 + \alpha_z 
(\partial_z \psi)^2 
\right]\right\},
\eea
with
$\psi(\rho=0,z)=0$ 
and $\psi(\rho,z=0)=\pi/2$ by symmetry. For notational simplicity, we have
suppressed the coordinate labels on $n,m,\psi$ in the above functional.
\begin{figure}
\includegraphics[width=3.2in]{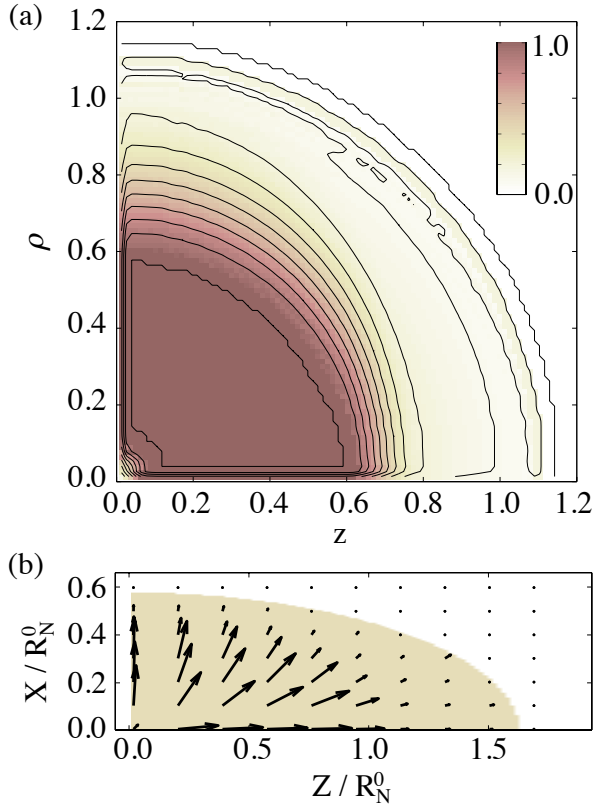}
\caption{(Color online)
Two-dimensional magnetisation profile for the distorted hedgehog showing 
breakdown of the LDA for the magnetization density for $10^2$ atoms in an
anisotropic trap with $\lambda=2.4$, and
$\alpha_{\perp} = 2, \alpha_{z} = 0.25$ ($\omega_\perp/\omega_z\approx 2.8$).
(a) Plot of the magnitude of the magnetization $m(\rho,z)$ and equal-magnetization
contours displayed in
rescaled coordinates in which the trap potential is spherically symmetric. Colorbar to the 
right indicates the value of $m(\rho, z)$.  We see that $m(\rho,z)$ is
larger in magnitude for larger values of $\rho$ than it is for $z$, indicating that
the surface tension makes it easier to change its value in the weak trapping direction. 
(b) $\vec m$ shown as a quiver plot indicating the magnitude and
direction of the magnetization (plotted in coordinates where the trap anisotropy
is explicitly shown). 
Shaded area indicates the region of the trap where the atom density 
is nonzero.}
\label{fig:aniso}
\end{figure}

We find, numerically, that $\psi \approx \theta$, so in fact the ansatz simplifies to the
form 
\be
\vec{m} ({\bf r}) = m(\rho,z) \left(\frac{x}{r},\frac{y}{r},\frac{z}{r}\right).
\ee
The main
effect of going from the spherical to the cigar shaped trap is that the magnitude of the
magnetization is no longer just dependent on the radial coordinate $r$. The magnetization
however still points (in our  rescaled coordinates) along the radial direction. The plot of
the magnetization for $\lambda=2.4$
in the rescaled and in the original coordinates for a trap anisotropy
corresponding to $\alpha_\perp = 2,\alpha_z= 0.25$ (a trap 
frequency ratio $\omega_\perp/\omega_z\approx 2.8$) is given in 
Fig.~\ref{fig:aniso}.

\section{Effect of dipolar interactions}
Our results for the spin texture energetics and magnetization profiles have 
been obtained
by neglecting the role of the long range magnetic dipole interaction
between the fermions. The
dipole interaction will add to the magnetic energy of atoms in the trap.
In addition, it
will lead to spatial variations of the magnetic field seen by atoms 
within the trap and, thus, cause tend to cause
dephasing as atoms in different regions will precess at
different rates. Such effects are 
known to be important in $^{87}$Rb spinor Bose
condensates \cite{vengalattore,demler08}.
In order to estimate the dipole interaction energy and the
timescale of this dephasing, we have considered the
spatial variations of the dipole field for the simple case of the 
spherical trap. 

The expression
for the precession frequency at distance $r$ from the center of the 
spherical trap is, for the hedgehog state,
\bea
\nu^{dip}_{HH}(r)\!\! &\!\!=\!\!&\!\! \frac{1}{h} \frac{\mu_0}{4\pi} \mu_B^2 (\frac{2\sqrt{2}}{\pi\sqrt{3}}) 
(\frac{\hbar}{M \Omega})^{-3/2} N^{1/2} \nonumber \\
\!\!\!\!\!&\!\!\times\!\!&\!\!\! \int \!\! \frac{dr_1 d\theta ~ r^2_1 \sin\theta ~ m(r_1) n(r_1)}{(r^2+r^2_1-2 r r_1 \cos\theta)^{3/2}} F(r,r_1,\theta),\\
\!\!F(r,r_1,\theta)\!\!&\!\!=\!\!&\!\!\left[\cos\theta\! -\! 3\frac{(r_1-r\cos\theta)(r_1\cos\theta-r)}{(r^2+r^2_1-2 r r_1 \cos\theta)}
\right],
\eea
where $\mu_B \approx 9.27 \times 10^{-24} J/T$
is the Bohr magneton, and $\mu_0 = 4\pi \times 10^{-7} N/A^2$ is the
permeability of free space.
Evaluating this, we find that the typical value of (and also the variation in)
the precession frequency for 
$\lambda=2.5$, $N=10^6$,
and $\Omega=2\pi(170 {\rm Hz})$, is $\nu^{dip}_{HH} (^6{\rm Li}) 
\approx 0.03 {\rm Hz}$ and 
$\nu^{dip}_{HH}(^{40}{\rm K})
\approx 0.6 {\rm Hz}$. The energy associated with the
dipole interactions is far smaller than our estimate of magnetic exchange energies,
$\sim 500$Hz,
arising from the s-wave contact interaction between fermions (in the interaction
range where we expect ferromagnetism). At the same time,
measurements of the typical atom lifetime, $\tau$, on the 
repulsive side of the Feshbach resonance indicate that $\tau \sim 10$ms for
$^{40}$K \cite{regal04} and $\tau\sim 100$ms for 
$^{6}$Li \cite{dieckmann02, bourdel03}. These are clearly much less than the 
variations 
in the precession period induced by spatial variations of the dipolar field
as estimated above. Taken together, these estimates show that ignoring the effect of dipole
interactions is a very good approximation in this system.

\section{Conclusions}

In conclusion, we have studied ferromagnetism and spin textures in 
ultracold atomic Fermi
gases in the regime of strongly repulsive interactions using the LDA
extended to include magnetization gradient corrections.  
Within the LDA at zero temperature,
we have shown that the release energy of the gas, as well as its separate kinetic energy
and interaction energy components, shows a sharp signature of the ferromagnetic
transition. We have also shown that
the atom loss rate via three-body collisions has a peak very 
close to the ferromagnetic transition and it provides yet another diagnostic 
of the transition into the ferromagnetic state.
We have gone beyond the LDA by deriving a surface tension 
correction to the energy functional, which depends on atom number and the
trap-geometry,
and used it to study the energetics of various
spin textures in a two-component trapped Fermi gas. For a spherically symmetric trap, 
we find that
a hedgehog magnetization profile has lower energy than a domain-wall state.
For large atom numbers, the small surface tension leads to a
small energy difference
between the two spin textures and the results are close to those of the
LDA. In this
case, the surface tension is responsible for selecting the hedgehog 
state as having the lowest energy but
we have checked that it does not significantly change our results for
the release energy and the atom loss rates.
These continue to be useful, albeit indirect, 
diagnostics of the transition into the ferromagnetic state.
For elongated clouds, we have shown that
the surface tension term distorts the hedgehog states, in rescaled
coordinates where the trap is isotropic, in such a manner that
the magnitude of the magnetization changes more easily in the 
weak direction of the trap than would be expected on the basis of the LDA.
Such a breakdown of the LDA is more apparent for smaller atom numbers.
Finally, we have considered the effect of magnetic dipolar interactions 
on our results and
find that it is a 
good approximation to
ignore dipole interactions in this system. 

The typical atom loss rate
near the Feshbach resonance sets a constraint that the formation time
for the ferromagnetic state will have to be
on the order of tens of milliseconds for
$^{40}$K \cite{regal04}, and hundreds of milliseconds for
$^{6}$Li \cite{dieckmann02, bourdel03}, in order for it to be observed.
A direct way to probe the spin textures discussed here
would be through high resolution {\it in situ} magnetometry
as has been done for spinor Bose condensates
\cite{stamperkurn06}.
An experimental observation of 
ferromagnetism in
trapped Fermi gases would provide 
impetus for future theoretical work on finite temperature 
effects and collective modes in the strongly interacting regime.
\bigskip

\acknowledgments
We thank Rembert Duine, Gyu-Boong Jo, Wolfgang Ketterle, and Allan MacDonald for 
useful discussions. This work was supported by 
NSERC of Canada (JHT, AP, AAB), the Canadian Institute for Advanced
Research (JHT, LL). AP acknowledges support from
the Sloan Foundation, the Connaught Foundation, and the Ontario ERA.
JHT thanks the 
MIT-Harvard Center for Ultracold Atoms for hospitality during the preparation
of this manuscript.
\bigskip

\rule{8cm}{0.3mm}

\bibitem{salasnich00}{
L. Salasnich, B. Pozzi, A. Parola and L. Reatto, J. Phys. B.: At. Mol. Opt. Phys.
{\bf 33}, 3943 (2000)}
\bibitem{sogo02}{
T. Sogo and H. Yabu, \pra {\bf 66}, 043611 (2002)}
\bibitem{duine05}{
R. A. Duine and A. H. Macdonald, \prl {\bf 95}, 230403 (2005)}
\bibitem{tasaki98}{
See, for instance, H. Tasaki, Prog. Theor. Phys. {\bf 99}, 489 (1998)}
\bibitem{congjun08}{
S. Zhang, H.-H. Hung, and C. Wu, arXiv:0805.3031 (unpublished)}
\bibitem{coleman08}{
I. Berdnikov, P. Coleman, and S. H. Simon, arXiv:0805.3693 (unpublished)}
\bibitem{emueller}{
T. N. De Silva and E. J. Mueller, \prl {\bf 97}, 070402 (2006); 
S. S. Natu and
E. J. Mueller, arXiv.0802.2083 (unpublished); S. K. Baur, S. Basu, 
T. N. De Silva, and E. J. Mueller, arXiv:0901.2945 (unpublished).}
\bibitem{sensarma07}{
R. Sensarma, W. Schneider, R. B. Diener, and M. Randeria, arXiv:0706.1741
(unpublished).}
\bibitem{hulet06}{
G. B. Partridge, W. Li, Y. A. Liao, R. G. Hulet, M. Haque, and 
H. T. Stoof, \prl {\bf 97}, 190407 (2006)}
\bibitem{bourdel03}{
T. Bourdel, J. Cubizolles, L. Khaykovich, K. M. Magalh\"aes, S. J. Kokkelmans, 
G. V. Shlyapnikov, and C. Salomon, \prl {\bf 91}, 020402 (2003)}
\bibitem{gupta03}{
S. Gupta, Z. Hadzibabic, M. W. Zwierlein, C. A. Stan, K. Dieckmann, C. H. Schunk,
E. G. M. van Kempen, B. J. Verhaar, and W. Ketterle, Science {\bf 300}, 1723 (2003)}
\bibitem{stamperkurn06}{
L. E. Sadler, J. M. Higbie, S. R. Leslie, M. Vengalattore, and D. M. Stamper-Kurn,
Nature {\bf 443}, 312 (2006)}
\bibitem{zweirlein05}{
M. W. Zwierlein, J. R. Abo-Shaeer, A. Schirotzek, C. H. Schunck and W. Ketterle,  Nature {\bf 435}, 
1047 (2005)}
\bibitem{greiner03}{
M. Greiner, C. A. Regal and D. S. Jin, Nature {\bf 426}, 537 (2003)}
\bibitem{regal04}{
C. A. Regal, M. Greiner and D. S. Jin, \prl {\bf 92}, 083201 (2004)}
\bibitem{dieckmann02}{
K. Dieckmann, C. A. Stan, S. Gupta, Z. Hadzibabic, C. H. Schunck, and W. Ketterle,  \prl {\bf 89}, 203201 (2002)}
\bibitem{stoner33}
{E. C. Stoner,  Phil. Mag. {\bf 15}, 1018 (1933)}
\bibitem{demler06}{
A. Imambekov, C. J. Bolech, M. Lukin, E. Demler, \pra {\bf 74}, 053626 (2006)}
\bibitem{vengalattore}{
M. Vengalattore, S. R. Leslie, J. Guzman, and D. M. Stamper-Kurn, \prl {\bf 100}, 
170403 (2008);
M. Vengalattore, J. Guzman, S. Leslie, F. Serwane and D. M. Stamper-Kurn, arXiv:0901.3800
(unpublished)}
\bibitem{demler08}{
R. W. Cherng and E. Demler, arXiv:0806.1991 (unpublished)}
\bibitem{dpetrov03}{
D. Petrov, \pra {\bf 67}, 010703 (2003)}
\listrefsAIP
\end{document}